\documentclass[aps, prd, twocolumn, reprint, preprintnumbers, floatfix, nofootinbib, superscriptaddress, 10pt]{revtex4-1}

\usepackage{graphicx}
\usepackage{enumitem}

\usepackage{amsmath}
\usepackage[table, svgnames, dvipsnames]{xcolor}
\usepackage{lineno}
\usepackage{hyperref}
\usepackage[normalem]{ulem} 
\usepackage{comment}
\begin{document}
\preprint{FERMILAB-PUB-25-0928-PPD}

\newcommand{\papertitle}{Ionization-based search for magnetic monopoles using the NOvA Far Detector}
\title{\papertitle}
\newcommand{\ANL}{Argonne National Laboratory, Argonne, Illinois 60439, 
USA}
\newcommand{\Bandirma}{Bandirma Onyedi Eyl\"ul University, Faculty of 
Engineering and Natural Sciences, Engineering Sciences Department, 
10200, Bandirma, Balıkesir, Turkey}
\newcommand{\ICS}{Institute of Computer Science, The Czech 
Academy of Sciences, 
182 07 Prague, Czech Republic}
\newcommand{\IOP}{Institute of Physics, The Czech 
Academy of Sciences, 
182 21 Prague, Czech Republic}
\newcommand{\Atlantico}{Universidad del Atlantico,
Carrera 30 No.\ 8-49, Puerto Colombia, Atlantico, Colombia}
\newcommand{\BHU}{Department of Physics, Institute of Science, Banaras 
Hindu University, Varanasi, 221 005, India}
\newcommand{\UCLA}{Physics and Astronomy Department, UCLA, Box 951547, Los 
Angeles, California 90095-1547, USA}
\newcommand{\Caltech}{California Institute of 
Technology, Pasadena, California 91125, USA}
\newcommand{\Cochin}{Department of Physics, Cochin University
of Science and Technology, Kochi 682 022, India}
\newcommand{\Charles}
{Charles University, Faculty of Mathematics and Physics,
 Institute of Particle and Nuclear Physics, Prague, Czech Republic}
\newcommand{\Cincinnati}{Department of Physics, University of Cincinnati, 
Cincinnati, Ohio 45221, USA}
\newcommand{\CSU}{Department of Physics, Colorado 
State University, Fort Collins, CO 80523-1875, USA}
\newcommand{\CTU}{Czech Technical University in Prague,
Brehova 7, 115 19 Prague 1, Czech Republic}
\newcommand{\Dallas}{Physics Department, University of Texas at Dallas,
800 W. Campbell Rd. Richardson, Texas 75083-0688, USA}
\newcommand{\DallasU}{University of Dallas, 1845 E 
Northgate Drive, Irving, Texas 75062 USA}
\newcommand{\Delhi}{Department of Physics and Astrophysics, University of 
Delhi, Delhi 110007, India}
\newcommand{\JINR}{Joint Institute for Nuclear Research,  
Dubna, Moscow region 141980, Russia}
\newcommand{\Erciyes}{
Department of Physics, Erciyes University, Kayseri 38030, Turkey}
\newcommand{\FNAL}{Fermi National Accelerator Laboratory, Batavia, 
Illinois 60510, USA}
\newcommand{\FSU}{Florida State University, Tallahassee, Florida 32306, USA}
\newcommand{\UFG}{Instituto de F\'{i}sica, Universidade Federal de 
Goi\'{a}s, Goi\^{a}nia, Goi\'{a}s, 74690-900, Brazil}
\newcommand{\Guwahati}{Department of Physics, IIT Guwahati, Guwahati, 781 
039, India}
\newcommand{\Harvard}{Department of Physics, Harvard University, 
Cambridge, Massachusetts 02138, USA}
\newcommand{\Houston}{Department of Physics, 
University of Houston, Houston, Texas 77204, USA}
\newcommand{\IHyderabad}{Department of Physics, IIT Hyderabad, Hyderabad, 
502 205, India}
\newcommand{\Hyderabad}{School of Physics, University of Hyderabad, 
Hyderabad, 500 046, India}
\newcommand{\IIT}{Illinois Institute of Technology,
Chicago IL 60616, USA}
\newcommand{\Indiana}{Indiana University, Bloomington, Indiana 47405, 
USA}
\newcommand{\INR}{Institute for Nuclear Research of Russia, Academy of 
Sciences 7a, 60th October Anniversary prospect, Moscow 117312, Russia}
\newcommand{\UIowa}{Department of Physics and Astronomy, University of Iowa, 
Iowa City, Iowa 52242, USA}
\newcommand{\ISU}{Department of Physics and Astronomy, Iowa State 
University, Ames, Iowa 50011, USA}
\newcommand{\Irvine}{Department of Physics and Astronomy, 
University of California at Irvine, Irvine, California 92697, USA}
\newcommand{\Jammu}{Department of Physics and Electronics, University of 
Jammu, Jammu Tawi, 180 006, Jammu and Kashmir, India}
\newcommand{\Lebedev}{Nuclear Physics and Astrophysics Division, Lebedev 
Physical 
Institute, Leninsky Prospect 53, 119991 Moscow, Russia}
\newcommand{\Magdalena}{Universidad del Magdalena, Carrera 32 No 22-08 Santa Marta, Colombia}
\newcommand{\MSU}{Department of Physics and Astronomy, Michigan State 
University, East Lansing, Michigan 48824, USA}
\newcommand{\Crookston}{Math, Science and Technology Department, University 
of Minnesota Crookston, Crookston, Minnesota 56716, USA}
\newcommand{\Duluth}{Department of Physics and Astronomy, 
University of Minnesota Duluth, Duluth, Minnesota 55812, USA}
\newcommand{\Minnesota}{School of Physics and Astronomy, University of 
Minnesota Twin Cities, Minneapolis, Minnesota 55455, USA}
\newcommand{\Mississippi}{University of Mississippi, University, Mississippi 38677, USA}
\newcommand{\NISER}{National Institute of Science Education and Research, An OCC of Homi Bhabha
National Institute, Bhubaneswar, Odisha, India}
\newcommand{\OSU}{Department of Physics, Ohio State University, Columbus,
Ohio 43210, USA}
\newcommand{\Oxford}{Subdepartment of Particle Physics, 
University of Oxford, Oxford OX1 3RH, United Kingdom}
\newcommand{\Panjab}{Department of Physics, Panjab University, 
Chandigarh, 160 014, India}
\newcommand{\Pitt}{Department of Physics, 
University of Pittsburgh, Pittsburgh, Pennsylvania 15260, USA}
\newcommand{\QMU}{Particle Physics Research Centre, 
Department of Physics and Astronomy,
Queen Mary University of London,
London E1 4NS, United Kingdom}
\newcommand{\RAL}{Rutherford Appleton Laboratory, Science 
and 
Technology Facilities Council, Didcot, OX11 0QX, United Kingdom}
\newcommand{\SAlabama}{Department of Physics, University of 
South Alabama, Mobile, Alabama 36688, USA} 
\newcommand{\Carolina}{Department of Physics and Astronomy, University of 
South Carolina, Columbia, South Carolina 29208, USA}
\newcommand{\SDakota}{South Dakota School of Mines and Technology, Rapid 
City, South Dakota 57701, USA}
\newcommand{\SMU}{Department of Physics, Southern Methodist University, 
Dallas, Texas 75275, USA}
\newcommand{\Stanford}{Department of Physics, Stanford University, 
Stanford, California 94305, USA}
\newcommand{\Sussex}{Department of Physics and Astronomy, University of 
Sussex, Falmer, Brighton BN1 9QH, United Kingdom}
\newcommand{\Syracuse}{Department of Physics, Syracuse University,
Syracuse NY 13210, USA}
\newcommand{\Tennessee}{Department of Physics and Astronomy, 
University of Tennessee, Knoxville, Tennessee 37996, USA}
\newcommand{\Texas}{Department of Physics, University of Texas at Austin, 
Austin, Texas 78712, USA}
\newcommand{\Tufts}{Department of Physics and Astronomy, Tufts University, Medford, 
Massachusetts 02155, USA}
\newcommand{\UCL}{Physics and Astronomy Department, University College 
London, 
Gower Street, London WC1E 6BT, United Kingdom}
\newcommand{\Virginia}{Department of Physics, University of Virginia, 
Charlottesville, Virginia 22904, USA}
\newcommand{\WSU}{Department of Mathematics, Statistics, and Physics,
 Wichita State University, 
Wichita, Kansas 67260, USA}
\newcommand{\WandM}{Department of Physics, William \& Mary, 
Williamsburg, Virginia 23187, USA}
\newcommand{\Wisconsin}{Department of Physics, University of 
Wisconsin-Madison, Madison, Wisconsin 53706, USA}
\newcommand{\deceased}{Deceased.}
\affiliation{\ANL}
\affiliation{\Atlantico}
\affiliation{\Bandirma}
\affiliation{\BHU}
\affiliation{\Caltech}
\affiliation{\Charles}
\affiliation{\Cincinnati}
\affiliation{\Cochin}
\affiliation{\CSU}
\affiliation{\CTU}
\affiliation{\Delhi}
\affiliation{\Erciyes}
\affiliation{\FNAL}
\affiliation{\FSU}
\affiliation{\UFG}
\affiliation{\Guwahati}
\affiliation{\Houston}
\affiliation{\Hyderabad}
\affiliation{\IHyderabad}
\affiliation{\IIT}
\affiliation{\Indiana}
\affiliation{\ICS}
\affiliation{\INR}
\affiliation{\IOP}
\affiliation{\UIowa}
\affiliation{\ISU}
\affiliation{\Irvine}
\affiliation{\JINR}
\affiliation{\Magdalena}
\affiliation{\MSU}
\affiliation{\Duluth}
\affiliation{\Minnesota}
\affiliation{\Mississippi}
\affiliation{\OSU}
\affiliation{\NISER}
\affiliation{\Panjab}
\affiliation{\Pitt}
\affiliation{\QMU}
\affiliation{\SAlabama}
\affiliation{\Carolina}
\affiliation{\SMU}
\affiliation{\Sussex}
\affiliation{\Syracuse}
\affiliation{\Texas}
\affiliation{\Tufts}
\affiliation{\UCL}
\affiliation{\Virginia}
\affiliation{\WSU}
\affiliation{\WandM}
\affiliation{\Wisconsin}

\author{S.~Abubakar}
\affiliation{\Erciyes}

\author{M.~A.~Acero}
\affiliation{\Atlantico}

\author{B.~Acharya}
\affiliation{\Mississippi}

\author{P.~Adamson}
\affiliation{\FNAL}









\author{N.~Anfimov}
\affiliation{\JINR}


\author{A.~Antoshkin}
\affiliation{\JINR}


\author{E.~Arrieta-Diaz}
\affiliation{\Magdalena}

\author{L.~Asquith}
\affiliation{\Sussex}


\author{A.~Aurisano}
\affiliation{\Cincinnati}



\author{A.~Back}
\affiliation{\Indiana}
\affiliation{\ISU}



% ...

\collaboration{The NOvA Collaboration}
\noaffiliation




\date{\today}

\begin{abstract}

We report a search for highly-ionizing magnetic monopoles in the cosmic-ray flux using a 2,713-day dataset collected during 2015--2025 with the NOvA Far Detector, a 14-kiloton segmented detector located on the Earth's surface in Minnesota, United States. The search is sensitive to monopoles across a wide range of speeds, \mbox{$7 \times 10^{-4} < \beta < 0.995$}, and is sensitive to masses as low as \mbox{$2 \times 10^5~\mathrm{GeV}$} for the fastest monopoles. No signal was observed. With the detector's large surface area and minimal overburden, we achieve the strongest flux limits reported to date in several regions of speed and mass. For heavy monopoles with masses above $10^{13}$~GeV that are able to reach the detector from above or --- crossing the Earth --- from below, we find a flux limit \mbox{$\phi_{90\%} < 2 \times 10^{-16}\,  \mathrm{ cm^{-2} s^{-1} sr^{-1}}$} (90\% C.L.) for monopoles with $0.005 < \beta < 0.8$. Across the same range of speeds, we report a limit \mbox{${\phi_{90\%}} < 8 \times 10^{-16}\, \mathrm{ cm^{-2} s^{-1} sr^{-1}}$} for light monopoles with masses above $10^8$~GeV that can reach the detector from above.

\end{abstract}

\maketitle


\section{Introduction}

Magnetic monopoles remain one of the most compelling and elusive mysteries in modern physics. Despite strong theoretical motivations and decades of experimental effort, not a single particle with a magnetic charge has ever been observed. The existence of a single magnetic pole, proposed by Paul Dirac in 1931 through the quantization of electric charge, is known as Dirac's quantization condition~\cite{Diracmonopole}. Dirac’s proposal suggests that even one magnetic monopole in the universe would explain the observed quantization of electric charge. According to the quantization condition, the magnetic charge {\it g} is an integral multiple of \(\frac{e}{2\alpha}\), where $e$ is the fundamental electric charge and $\alpha$ is the fine structure constant. Under this hypothesis, the ionization strength of the smallest possible magnetic charge, known as the Dirac charge $g_D$, is equivalent to that of a particle carrying  68.5 times the fundamental electric charge.

Following Dirac's insight, certain frameworks within string theory suggest that the requirement of electric charge quantization implies the existence of magnetic monopoles~\cite{josephP}. The existence of even a single monopole would restore the symmetry between electric and magnetic fields in Maxwell's equations. 
Monopoles are allowed in Grand Unified Theories (GUT) as fundamental solutions to spontaneous symmetry breaking and the associated topology of the unified gauge field~\cite{gutPDG, gutHooft, gutPolyakov}. They are created as stable topological defects when the unified gauge symmetry breaks down into fundamental forces. They are extremely massive, with masses close to the unification scale, typically higher than $10^{16}~\mathrm{GeV}$, and carry a magnetic charge quantized in units of the Dirac charge. Some theoretical models also predict magnetic monopoles with masses as low as $10^7~\mathrm{GeV}$~\cite{Kephart}. All of these GUT monopoles are massive compared to Standard Model particles and cannot be produced in existing or foreseen particle accelerators. In addition to superheavy GUT monopoles, several other theories predict monopoles at the electroweak scale~\cite{ChoandMaison, Rajantie, Hung, cho2014masselectroweakmonopole}.

If the laws of physics allow magnetic monopoles to exist, GUT monopoles could have
formed in the early universe and still exist in the cosmic flux~\cite{Zeldovich, cosmicmonopoles}. Astrophysical considerations put an upper limit on the GUT monopole flux at $ 10^{-15}\,\mathrm{cm^{-2}\, s^{-1}\, {sr}^{-1}}$, a constraint known as the Parker bound~\cite{parkerbound}. These relic monopoles have been sought in the cosmic flux using various techniques. One of the early experimental efforts was conducted by Cabrera {\it{et al.}}, who used superconducting rings as monopole detectors~\cite{cabrera1, cabrera2}. Another approach involves probing for monopole-induced proton decay catalysis, as proposed in several theoretical works~\cite{Rubakov-protondecay1, Rubakov-protondecay2, Callan-protondecay3} and 
investigated experimentally by Super-Kamiokande~\cite{ProtondecaySK}. Mountaintop experiments,
such as SLIM, have employed nuclear track detectors~\cite{slimresult}. Underground 
scintillator-based experiments, such as MACRO, have presented the strongest limit on GUT monopoles with $g= g_D$ at the level $1.4 \times 10^{-16}~\mathrm{cm^{-2} 
s^{-1}sr^{-1}}$ for a very wide range of speeds: $4\times 10^{-5} < \beta < 1$~\cite{macroresult2}, where $\beta$ denotes the ratio of the monopole’s speed to the speed of light.

The IceCube~\cite{IceCube1, IceCube2} and ANTARES~\cite{antares, antaressearchmagneticmonopolescomplete} experiments, both employing the Cherenkov detection method, have searched for magnetic monopoles. IceCube, located in the Antarctic ice, has set the most stringent flux limits, reaching below $2 \times 10^{-19}~\mathrm{cm^{-2}\,s^{-1}\,sr^{-1}}$ for $0.8 \leq \beta \lesssim 0.995$~\cite{IceCube2}. In addition, searches have
been conducted on lunar rock samples, particularly in studies performed at 
Berkeley~\cite{berkeley1, berkeley2}. Monopoles have also been probed through their possible production in proton-proton collisions at the LHC~\cite{moedal1, moedal2, moedal3, atlas_prl, atlas_jhep} and production in collisions of cosmic rays bombarding the atmosphere~\cite{atmoprod}. The Far Detector of the NuMI Off-Axis $\nu_e$ Appearance (NOvA) experiment, located on the Earth's surface with minimal overburden and a large surface area, provides remarkable sensitivity for even low-mass monopoles, which are generally not accessible in underground experiments. 
 
 This paper is organized as follows. Section~II describes the NOvA detectors. Section~III provides a detailed description of the procedure used to simulate the signal sample to evaluate the efficiency of this analysis, as well as the method used to determine the accessible solid angle. Section~IV describes the online trigger that initiated the data collection used in this analysis, and Sec.~V presents the offline analysis strategy. Finally, the results are presented in Sec.~VI.

\section{NOvA Far Detector}
NOvA is a long-baseline neutrino experiment~\cite{novaTDR} that uses an intense beam of muon (anti)neutrinos and two highly segmented and functionally identical detectors to investigate key questions in the neutrino sector, such as the measurement of mixing parameters, the mass ordering, and the CP-violating phase \(\delta_{CP}\). The 0.3\,kt Near Detector is located 1\,km downstream of the neutrino production target and 100\,m underground at Fermilab, and the 14\,kt Far Detector (FD) is located on the surface, 810\,km from the target, near Ash River, Minnesota. The detectors are 14.6\,mrad off the axis of the NuMI beam, producing a narrow-band neutrino flux that peaks at 1.8\,GeV.

  \begin{figure}
   \centering
    \includegraphics[width=\columnwidth]{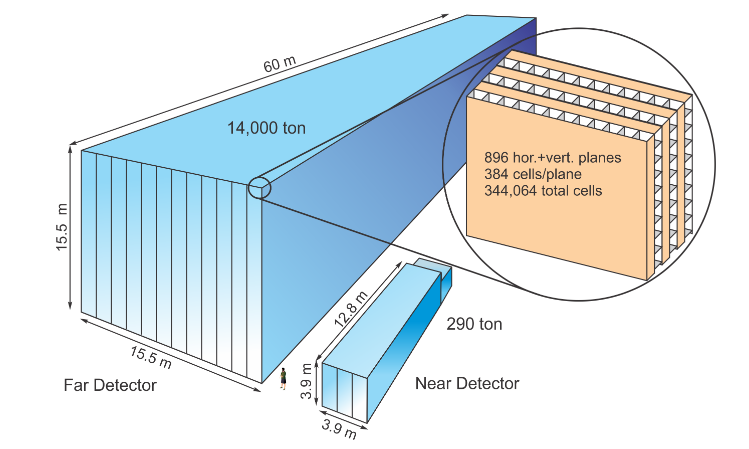}
   \caption{Schematic diagram of the NOvA detectors. For each detector, the coordinate system origin is set at the center of the front face. The {\it z}-axis points forward along the length of the detector, the {\it y}-axis points vertically upward, the {\it x}-axis points horizontally along the width of the detector, and together they form an orthogonal right-handed system. The inset illustrates the internal structure of the detectors, where cells are arranged in alternating vertical and horizontal orientations. This geometry enables precise 3D reconstruction of particle trajectories and energy deposition.}
   \label{fig:novadetectors}
\end{figure}

Shielded only by a 3.6\,m water equivalent overburden, the FD is exposed to a cosmic-ray flux of 150~kHz. It is a segmented tracking calorimeter, of dimensions $\mathrm{15.5~ m \times 15.5~ m \times 59.8~ m}$, with 896 planes of 384 PVC cells~\cite{novapvc}. The cells are filled with an organic scintillator (mineral oil mixed with 5\% pseudocumene), which accounts for 63\% of the mass of the detector~\cite{novascintillator}. The cells have dimensions {$\mathrm{1,550~cm \times 3.9~cm \times 6.6~cm}$} and are oriented in alternating vertical and horizontal views, as shown in Fig.~\ref{fig:novadetectors}, to achieve three-dimensional (3D) event reconstruction. Planes with horizontal cells make up the YZ view, while planes with vertical cells make up the XZ view. Each cell contains a looped wavelength-shifting (WLS) fiber that runs along its length to collect light, which is guided to a photodetector. NOvA uses avalanche photodiodes (APDs) as photodetectors. Both ends of the fiber from each cell are coupled to one pixel of a 32-pixel APD array. 

When a charged particle passes through a scintillator-filled cell, it produces light that is captured and converted from blue (400--450~nm) to green (490--550~nm) by the fluorescent dye in the WLS fiber. A portion of the light emitted by the dye is guided within the fiber via total internal reflection. As it travels along the fiber, a significant fraction of the shorter‑wavelength component ($<$ 520~nm) is absorbed, while longer‑wavelength light is only weakly attenuated. In the wavelength range corresponding to the fiber output, the APD exhibits a quantum efficiency of 85\% and produces a robust signal for the minimum ionizing particles passing through any point along the length of a cell. The APD converts light into an electrical signal, which is continuously sampled every 500~ns and digitized by an analog-to-digital converter (ADC). The resulting digital signal appears as a pulse, whose height is proportional to the energy deposited in the cell. In the following text, we use \emph{ADC} to refer specifically to the digitized signal output. A \emph{hit} is recorded in a cell when the digitized samples meet a predefined criterion set individually for each APD pixel. The thresholds are defined as three times the measured RMS noise at installation, with the APDs both cooled and biased. This corresponds to roughly one-third of a minimum ionizing particle (MIP) signal originating from the far end. Variations between individual pixels can be as large as 10\%. Due to the alternating orientation of detector planes, each hit provides two-dimensional (2D) position information from either the XZ or YZ view, along with the time and magnitude of the energy deposited. The signature of the passage of a magnetic monopole through FD is discussed in more detail in Sec.~\ref{sec:simulation}.

\section{Simulation}
\label{sec:simulation}

Simulation is used to determine two inputs to the analysis: first, the detection efficiency, which measures how well the FD can identify a monopole once it arrives; and second, the solid angle coverage, which gives the fraction of the sky over which a monopole of a given speed and mass can reach the FD.

To evaluate the signal selection efficiency, a sample of magnetic monopoles, isotropically distributed, each carrying a magnetic charge of $g_D$ and no electric charge, is generated using the standard NOvA simulation framework~\cite{novasimulationchain}. The energy loss of monopoles within the detector is modeled using GEANT4~\cite{geant4}, while light production, propagation, and detection are modeled using custom simulation code developed specifically for NOvA detectors~\cite{novasimulationchain}. Simulated monopoles with a velocity distribution in the range $10^{-4} \le \beta \le 0.995$ are uniformly distributed on all surfaces of the FD. The mass of the simulated monopoles is treated as heavy enough so that the energy loss of monopoles in the detector does not affect their speed. We do not consider the catalysis of proton decay by magnetic monopoles because it is predicted that this process does not occur at a significant rate~\cite{Rubakov-protondecay1, Callan-protondecay3}. If such a process occurs, NOvA's sensitivity to low-mass monopoles would be reduced, in an amount depending on the catalysis model, as monopoles would be lost in the Earth and in the atmosphere. 

Energy deposition of magnetic monopoles in the FD occurs primarily through ionization and atomic excitation. The stopping power formula proposed by Ahlen and Kinoshita~\cite{AhlenKinoshita} is implemented in our simulation for monopoles with $\beta < 10^{-2}$:
 \begin{equation}
    \frac{dE}{dx} = C_{a}\beta,
    \label{eq:dedx_slow}
\end{equation}
where 
\begin{equation}
    C_{a} = \frac{2\pi N_e g^2 e^2}{m_e v_F} \left[\ln \frac{2 m_e v_F a_0}{\hbar} - 0.5 \right] = 12~\mathrm{GeV/cm},
\end{equation}
where $v_F$ is the Fermi velocity $(\frac{\hbar}{m_e}) (3 \pi^2 N_e)^{\frac{1}{3}}$, $a_0$ is the Bohr radius and $N_e$ is the density of electrons. Similarly, the following stopping power model calculated by Ahlen~\cite{Ahlen} is implemented for monopoles with \mbox{$\beta > 0.1$}:

\begin{multline}
\frac{dE}{dx} = \frac{4\pi N_e g^2 e^2}{m_e}
\Bigg[ \ln \left( \frac{2 m_e \beta^2 \gamma^2}{I} \right)
- \frac{1 + \delta}{2} \\
 - B(|g|) + \frac{K(|g|)}{2}\Bigg],
\label{eq:dedx_fast}
\end{multline}
 where $B (| g |) = 0.248$ is the Bloch correction and $K (| g |) = 0.406$ is the QED correction for $| g | = 137e/2$. The quantity $I$ denotes the mean ionization potential of the medium, and $\delta$ is the density-effect correction that accounts for the polarization of the medium at high energies. For reference, the energy loss at $\beta = 0.1$, 0.5, and 0.9 is 3.17~GeV/cm, 5.55~GeV/cm and 7.23~GeV/cm, respectively, as compared to that of a MIP, $2 \times 10^{-3}$~GeV/cm. For the intermediate region, $0.01 \leq \beta \leq 0.1$, we interpolate linearly between the two models.

A dedicated test stand~\cite{novabirks_2020} was built to accurately determine the Birks constant, allowing the characterization of the light response of the NOvA scintillator to highly ionizing particles. In this setup, a neutron source was used to generate protons within a small sample of NOvA scintillator, and light sensors recorded the resulting scintillation signals. A timing system measured the neutron time-of-flight to the scintillator, enabling the proton energies to be determined. The scintillation light produced by these protons was then compared to the response from well-characterized gamma-ray sources to obtain a calibration. By varying the Birks constant in the light-response model until the calculated light output agreed with the observed signal, its value was determined to be $(1.155\pm0.065) \times 10^{-2}~\mathrm{g~cm^{-2}~MeV^{-2}}$. Birks' law limits the scintillation signal, which does not increase significantly above $\beta \approx 0.1$ despite the increase in energy loss. In the limit of high dE/dx the scintillation light is as bright as it would be for $0.07$~GeV/cm if there were no Birks suppression, which is equivalent to approximately $35$ times the signal produced by a MIP. The test stand probed dE/dx values of up to 1\,GeV/cm. As shown in the examples above, monopoles are predicted to have energy losses an order of magnitude higher than this at high beta.  However, as discussed below, Cherenkov light dominates in this region, making the details of Birks suppression subdominant.

NOvA is primarily designed to study the energy deposition of relativistic particles. In contrast, magnetic monopoles may traverse the detector at velocities well below the speed of light. To evaluate the detector response to such slow, highly ionizing signals, a separate test stand measurement was performed. In this setup, APDs read out by NOvA electronics were exposed to LED light pulses designed to mimic the signatures of monopoles. The pulse durations matched the time a monopole would take to cross a cell, and the intensities were tuned to reproduce the anticipated energy deposition of monopoles at various speeds. The measured response showed good agreement with the simulation of the detector electronics within systematic uncertainty of 10\% assigned to the test stand measurement. Following the approach presented in Ref.~\cite{slowmononova}, to account for the possibility of an overestimated detector response to slow signals, we conservatively reduce the expected light detection by 10\% for monopoles with $\beta < 0.01$. For monopoles with $\beta > 0.1$, a smaller reduction of $3\%$ is applied to account for theoretical uncertainties~\cite{Ahlen}. A smooth linear interpolation is performed in the intermediate range $0.01 \leq \beta \leq 0.1$.

The scintillator has a wavelength-dependent refractive index, which is 1.47 at 400\,nm, resulting in a Cherenkov threshold for magnetic monopoles of $\beta \approx 0.68$. According to the magnetic monopole equivalent of the Frank-Tamm formula~\cite{Frank1991}, a relativistic monopole emits Cherenkov light at a rate proportional to $(gn/e)^2$ times that of a particle with electric charge $e$ traveling at the same speed in a medium with refractive index $n$. Above the Cherenkov threshold, a magnetic monopole with magnetic charge $g_D$ is therefore expected to produce $\mathcal{O}(10^4)$ times more light than a typical cosmic muon in the FD. For monopoles with $\beta > 0.68$, the total light yield includes contributions from both ionization energy loss, reduced by Birks suppression, and Cherenkov radiation. The passage of a magnetic monopole through the FD is expected to produce a uniform light yield along its path. This distinguishes monopoles from the obvious background, i.e. ordinary cosmic rays, which can be either minimally ionizing and uniform (low-energy muons) or highly ionizing and non-uniform (high-energy muons, protons, gammas, etc.).

As the monopole speed increases, the spectrum of secondary electrons, known as delta rays, begins to include particles that can travel several centimeters, reaching into neighboring detector cells. This effect becomes significant at $\beta \approx 0.68$ and leads to a significant reduction in detection efficiency, which is further discussed in Sec.~\ref{sec:offline_analysis}. However, explicitly generating delta rays for high speed monopoles is computationally expensive. To model this behavior, we instead use a custom simulation which simulates intense ionization within a cylinder around each simulated monopole, without explicitly generating each delta ray. This approximation is well motivated, as either the monopole itself, or the large number of delta rays will saturate any detector channel within the affected region, with a relatively sharp cutoff at the maximum transverse range of the delta rays for any given monopole speed.

To realistically model the detector environment during a potential magnetic monopole passage, we combine each simulated monopole with 5\, ms of minimum biased FD data (i.e., data recorded and saved to permanent storage without any selection). This resulting sample consists of events that contain both the simulated monopole signal and typical detector activity. 

\begin{figure}
    \centering
    \includegraphics[width=\columnwidth]{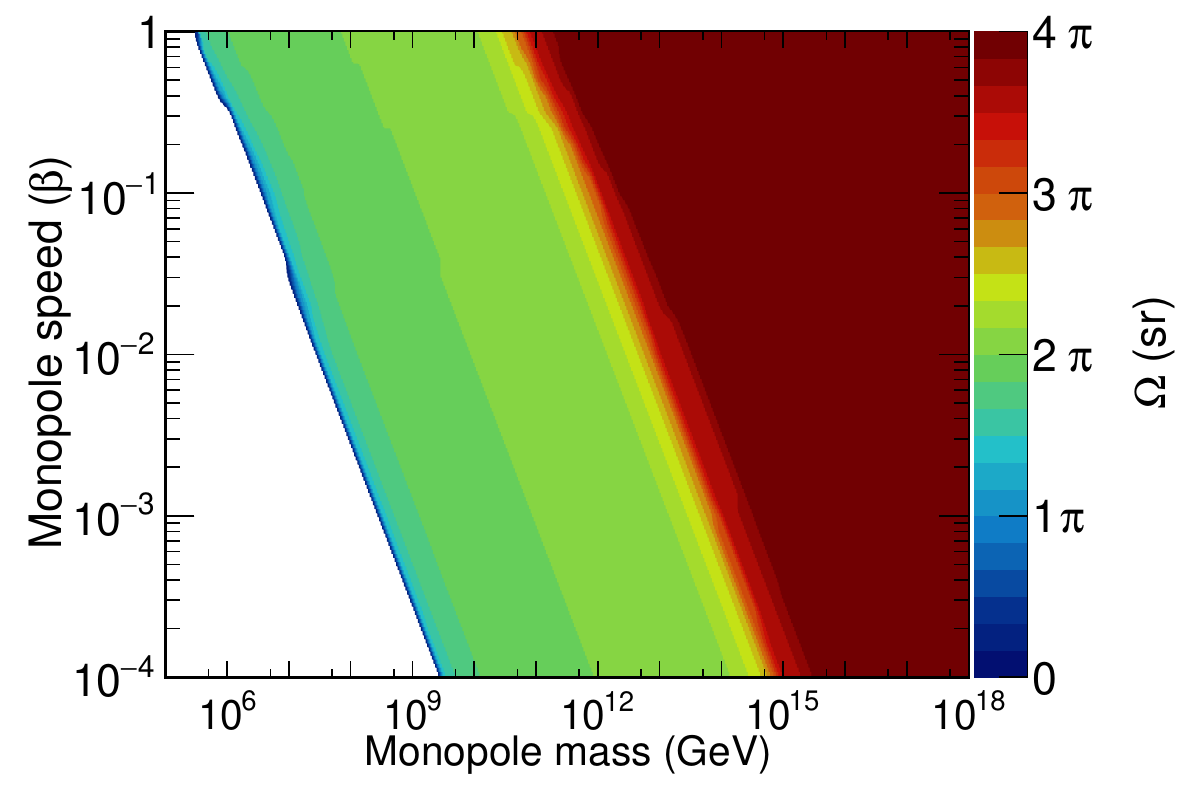}
    \includegraphics[width=\columnwidth]{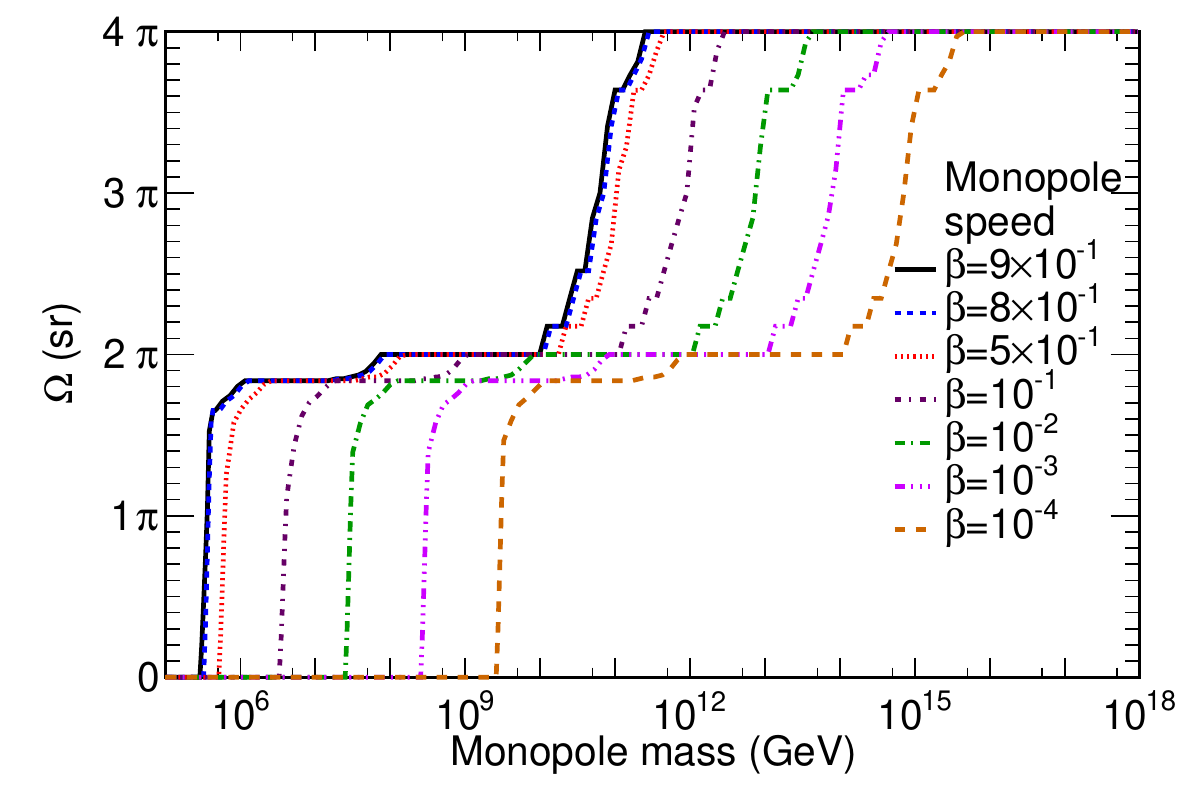}
    \caption{Solid angle coverage as a function of monopole speed ($\beta$) and mass (top), and as a function of monopole mass for selected values of monopole speed, $\beta = 10^{-4}$, $10^{-3}$, $10^{-2}$, $10^{-1}$, $5 \times 10^{-1}$, $8 \times 10^{-1}$ and  $9 \times 10^{-1}$ (bottom).}
\label{fig:solid-angle}
\end{figure} 
 
For each point in the space of monopole mass and speed, we used simulation to determine the fraction of isotropically distributed monopoles that can reach the detector. We denote this result as the \emph{accessible solid angle} $\Omega$. The calculation used the stopping powers from Eqs.~\ref{eq:dedx_slow} and \ref{eq:dedx_fast} in a geometry that included a model of the FD overburden and a simple three-layer earth model~\cite{Derkaouietal}. The resulting $\Omega(\beta,m)$ is shown in Fig.~\ref{fig:solid-angle}, with the top panel displaying the dependence on both $\beta$ and mass. For better visualization, the lower panel shows the coverage as a function of mass for selected values of $\beta$. A similar dependence of acceptance on monopole mass, for various magnetic charges and speeds, is shown in Ref.~\cite{Derkaouietal}; however, our results are specific to monopoles with a single Dirac charge ($g = g_D$).

In the limit of high monopole mass, the monopoles are able to traverse the entire Earth, which is an important special case that we refer to as the $4\pi$ region in Fig.~\ref{fig:solid-angle}. As monopole mass decreases, the accessible solid angle correspondingly decreases, with two main features: first, intermediate-mass monopoles can arrive only from above the horizon; and second, the lightest monopoles to which NOvA has sensitivity can arrive only through the thinnest part of the detector overburden, directly above the detector. In the lower panel  of Fig.~\ref{fig:solid-angle}, the point at which each curve drop to zero marks the lowest monopole mass that can be detected for the corresponding $\beta$.

\section{Online Trigger}
The NOvA FD uses several software-based data driven triggers~\cite{NovaDDTs}, including two designed specifically for magnetic monopoles: an ionization-based trigger and a transit time-based trigger. In NOvA, the ionization-based trigger records candidate events characterized by high energy deposition, whereas the transit time-based trigger identifies and records those with velocities well below the speed of light. The two categories are mainly for convenience in the trigger system and have some event overlap. In this work, we focus solely on the ionization-based trigger; results from the transit time-based trigger are presented elsewhere~\cite{slowmononova}. The ionization-based trigger continuously searches for long, through-going, highly ionizing tracks with uniform energy deposition against the background of 150~kHz of ordinary cosmic rays. A detailed analysis of the potential cosmic background relevant to the detection of magnetic monopoles in the FD is provided in Sec.~\ref{sec:offline_analysis}.

The first step of the trigger algorithm is to identify hits that are all associated with the same interaction. We call a group of hits correlated in space and time a \emph{slice}. Each slice has hits in the two views, XZ and YZ. Since monopoles are expected to deposit enormous energy in the detector, only hits with more than 500~ADC are clustered into slices. At the trigger level, we do not perform 3D tracking due to CPU time constraints, so absolute energy calibration with  attenuation correction is absent throughout the trigger decision-making process.

 The ionization-based trigger calculates a collection of variables from the slices that are useful for classification of interesting events. Since monopoles are expected to produce nearly uniform ionization along their path, the trigger algorithm looks for slices that are characterized by a high density of hits per unit length, small variations in the number of hits per plane, and an ADC-weighted center that remains close to the geometric center of the track. Since monopoles are expected to traverse the detector without stopping, the trigger also requires hits near opposite surfaces of the detector, consistent with the signature of a through-going particle. Additional conditions, such as a minimum number of surface hits and a minimum slice diagonal length, are applied to reject short or localized background events. Based on these requirements, the ionization-based trigger determines whether a slice contains a potential highly ionizing monopole track while maintaining a low overall trigger rate. If the criteria are satisfied, a trigger is issued with a time window equal to the duration of the slice, extended by a buffer of 781.25\,ns (50 clock ticks at 64~MHz) on each end. The resulting event is then saved to permanent storage for future analysis.

\begin{figure}
    \centering
    \includegraphics[width=\columnwidth]{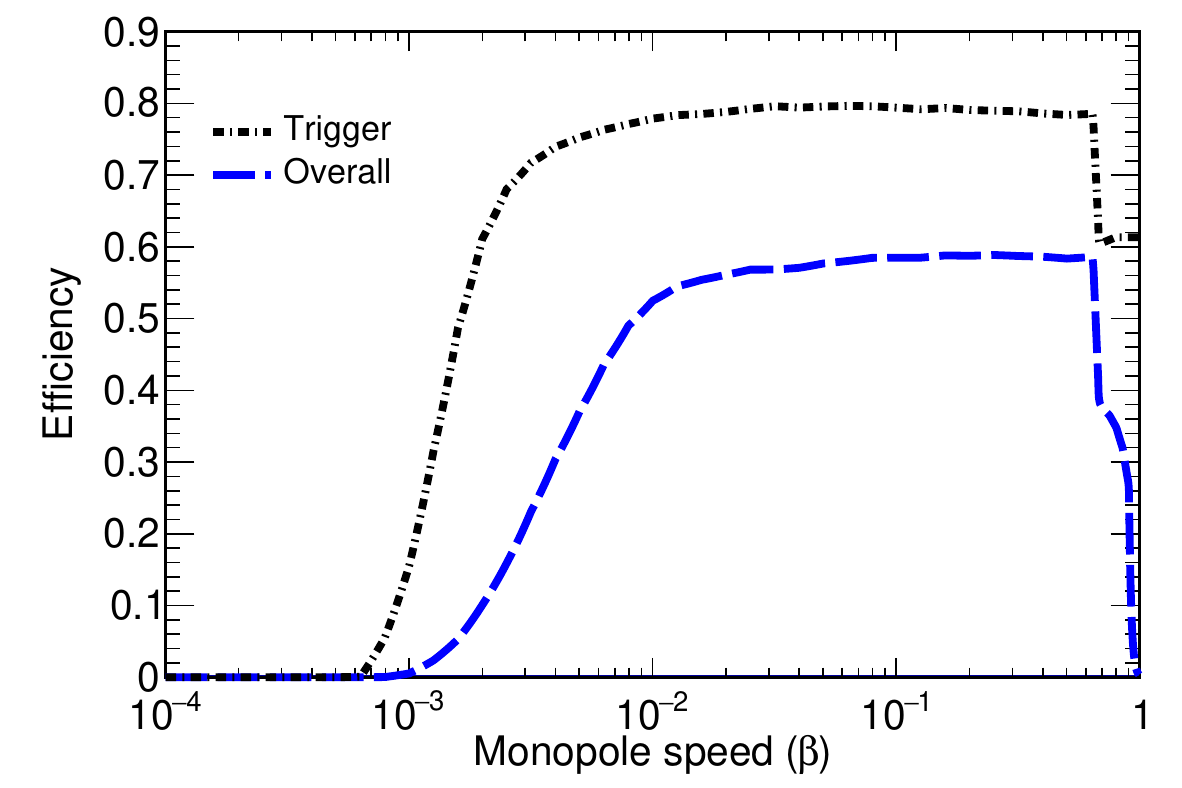}
    \caption{Trigger and overall efficiency as a function of monopole speed ($\beta$).} 
    \label{fig:trig_eff}
\end{figure}

Figure~\ref{fig:trig_eff} shows the \emph{trigger efficiency}, defined as the fraction of simulated monopoles, originating just outside the detector, that are selected by the trigger. The main reason for trigger inefficiency is slice reconstruction failure. Monopoles moving almost vertically through the detector have a greater chance to go through the gaps between two adjacent planes, thus leaving too few cell hits for slice reconstruction. Furthermore, for monopoles with $\beta < 0.01$, the energy deposited in each cell may not be enough to produce a hit or for a hit produced to exceed the ADC threshold required for slicing. For $\beta$ above the Cherenkov threshold, the efficiency decreases due to the crosstalk effect, as discussed in detail in Sec.~\ref{sec:offline_analysis}.

 \section{Offline Analysis}
 \label{sec:offline_analysis}
 Events that satisfy trigger requirements are further analyzed offline with more sophisticated reconstruction algorithms. In the offline analysis, 3D tracks are reconstructed to obtain several key variables for signal selection, such as velocity, track length, track width, etc. The process starts by grouping hits that are close in space and time. These clusters are used to reconstruct 2D tracks in the XZ and YZ views, which are then combined to form the full 3D track. 

 This search is based on data collected with the NOvA FD between November 2015 and February 2025. The data set corresponds to a total of 2,904~days of live time. After applying corrections for detector-specific data quality criteria and the fraction of time the data-driven trigger was operational, the effective good live time used in this analysis is calculated as 2,713~days.   

 The signal selection for this search is performed in two stages: preselection and final selection. The preselection stage applies relatively loose criteria to eliminate obvious background events while maintaining high signal efficiency. The final selection then imposes stricter requirements to completely suppress any residual background. To develop and optimize these selection strategies, we used a subset of ionization-based triggered data uniformly sampled from 2015 to 2019 and representing approximately 0.5\% of the total data set described above. Due to limited background statistics, the selection cuts were kept simple and not optimized with a formal figure of merit. Instead, the strategy was to reject all backgrounds while maximizing the \emph{overall selection efficiency}, defined as the fraction of simulated monopoles, originating just outside the detector, that pass the trigger, preselection, and final selection requirements.

 The preselection criteria are chosen on the basis of the expected properties of magnetic monopoles: highly ionizing, subluminal, through-going particles. Reconstructed tracks must have entry and exit points near the detector boundaries, sufficient hits in both XZ and YZ views with a long reconstructed length, and satisfy timing and velocity constraints to ensure subluminality. Furthermore, monopoles are expected to exhibit a large energy loss per unit length and a high average ADC per hit 
 (hereinafter called \emph{mean ADC}). The following preselection criteria are applied to each 3D reconstructed track: 
 
 \begin{enumerate}
  \item[1.] At least 20 hits in each view, with a reconstructed length of at least 5~m;
  \item[2.] Entry and exit points within 50~cm of the detector surface;
  \item[3.] Reconstructed velocity below the speed of light, and time difference between first and last hits of at least 350~ns;
  \item [4.] Track $dE/dx$ of at least $2.5 \times 10^{-3}$~GeV/cm, and mean ADC of at least 850.
  
\end{enumerate}

 More stringent cuts are applied in the final event selection to retain most monopole signal events while rejecting background tracks that pass the preselection. Given their large mass and momentum, monopoles are not expected to undergo significant multiple scattering, resulting in straight, narrow tracks within detector resolution. Events that passed the preselection criteria were filtered based on the track width, as discussed below, to eliminate backgrounds from Standard Model cosmic-ray tracks that can produce large, but irregular, energy depositions.

\begin{figure}
    \centering
    \includegraphics[width=\columnwidth]{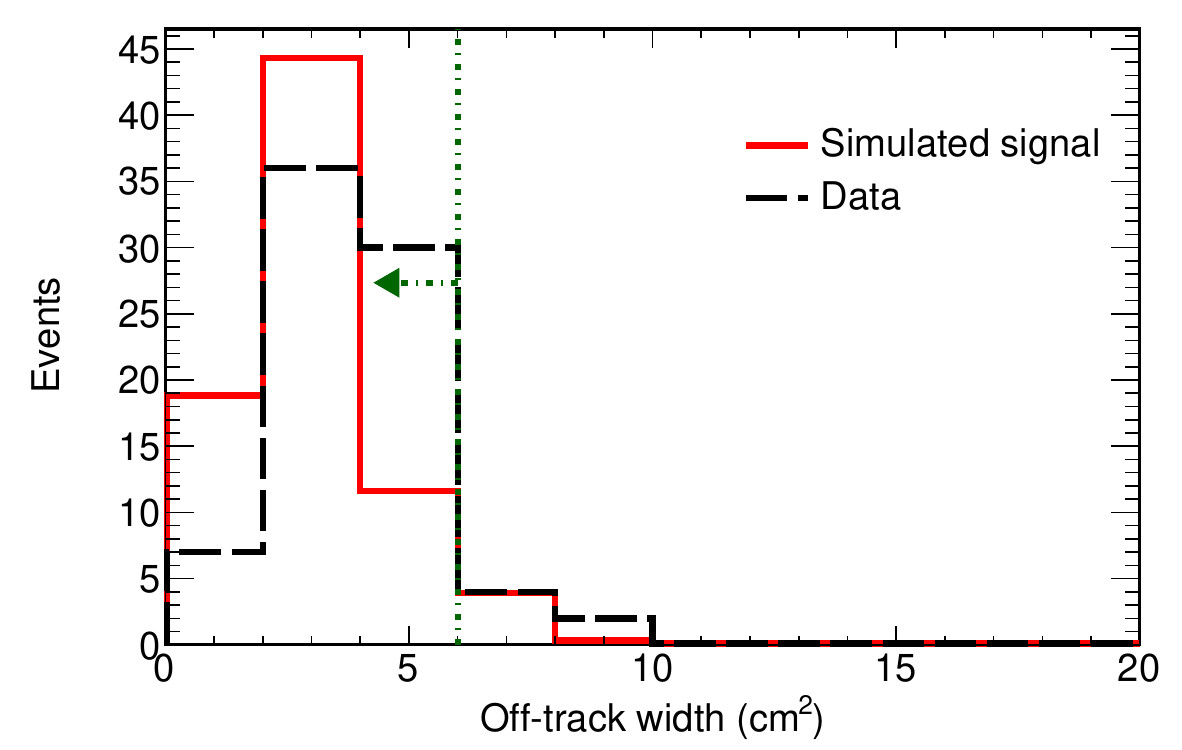}

    \includegraphics[width=\columnwidth]{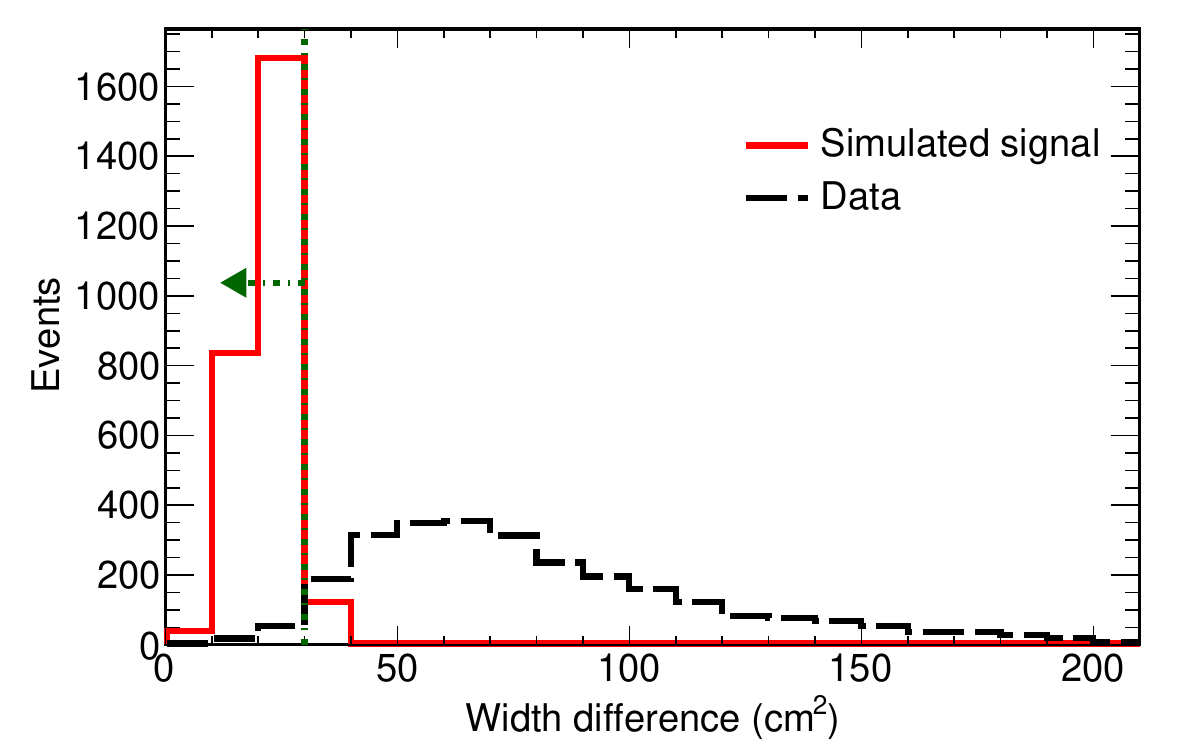}
    \caption{Distributions of off-track width (top) and width difference (bottom) for data and simulated signal sample. In each plot, the simulated event distribution is area-normalized to the data. The top (bottom) distribution corresponds to events passing the trigger, and selection requirements 1--4 and 6 (1--5). The dashed vertical line with an arrow indicates the cut value of the variable; note that this cut is not applied to the distribution shown.}
    \label{fig:sideband_N-1_trackwidth}
\end{figure}
The track width is defined as the average squared distance of the hits from the fitted track line, calculated by adding the squared distances of all hits and dividing by the number of hits. Two  variables related to track width are defined for signal selection:
 (1)~\emph{Off-track width}: the maximum value, taken between the XZ and YZ views, of the sum of squared distances from the reconstructed track for hits whose cells are not intersected by the straight-line reconstructed track; and (2)~\emph{width difference}: the maximum value, taken between the XZ and YZ views, of the difference between the maximum and minimum widths observed along the track.
 Figure \ref{fig:sideband_N-1_trackwidth} illustrates how these variables are used to separate monopoles from the Standard Model background. The track width selection requirement is the following:
 \begin{enumerate}
  \item[5.] Off-track width  $\leq 6\, \mathrm{cm}^2$; 
  \item[6.] Width difference $\leq 30\, \mathrm{cm}^2$.
\end{enumerate}

\begin{figure}
    \centering
     \includegraphics[width=\columnwidth]{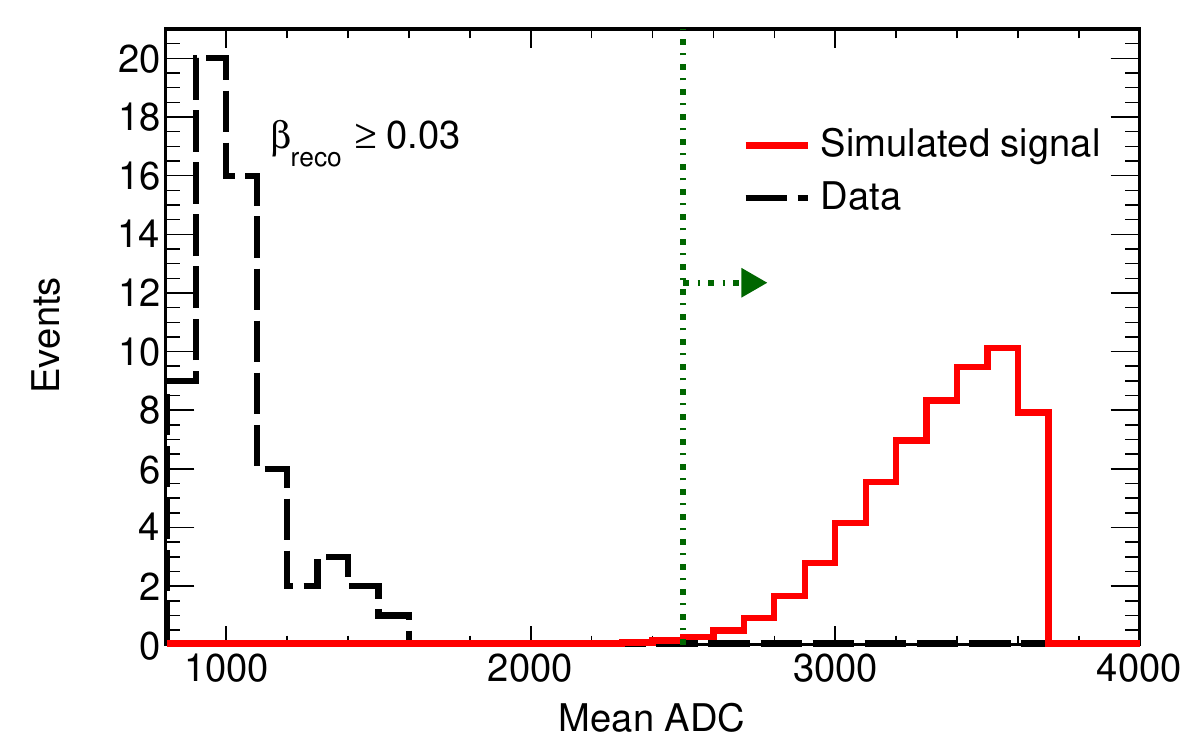}
    \includegraphics[width=\columnwidth]{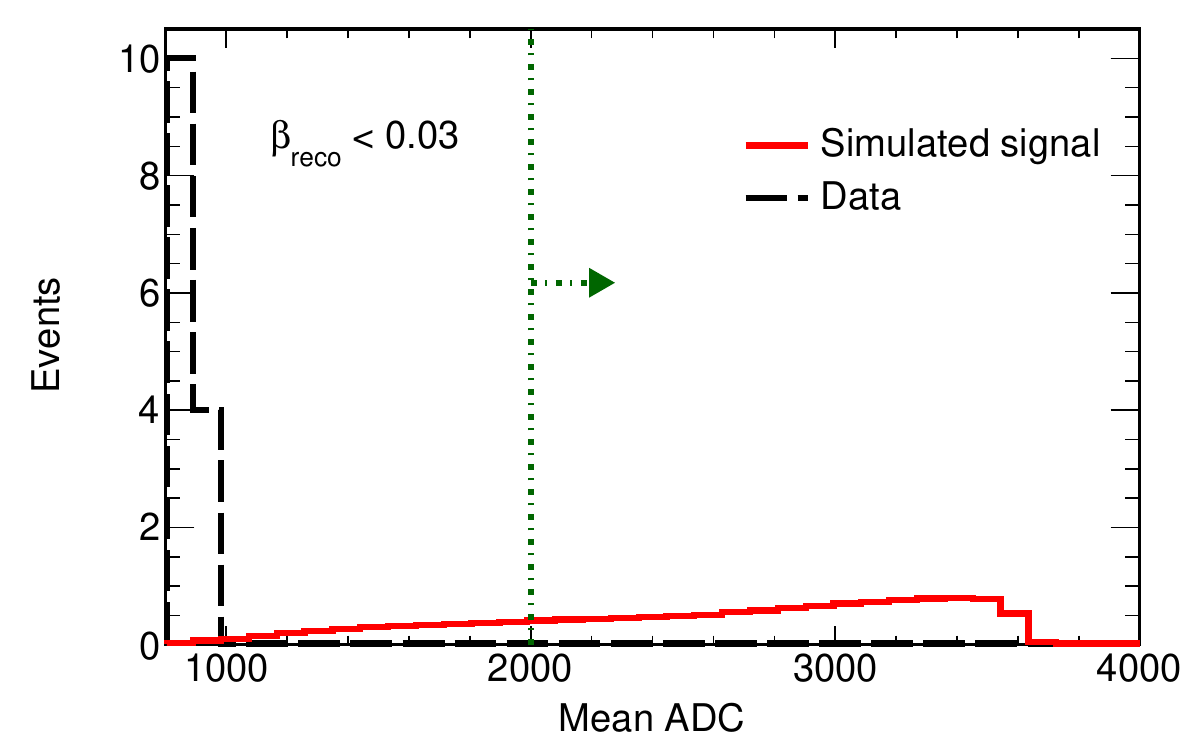}
    \caption{Distributions of mean ADC for data and simulated signal events, shown for events with $\beta_{\text{reco}} \geq 0.03$ (top) and $\beta_{\text{reco}} < 0.03$ (bottom). In each plot, the simulated event distribution is area-normalized to the data for visual comparison. In each case, trigger and selection requirements 1--6 have been applied. These plots are used to define the signal region (note suppressed zero on horizontal axis). The dashed vertical line with an arrow indicates the cut value of the variable; note that this cut is not applied to the distribution shown. Most of the background events are due to irregular or clustered cosmic activity and are rejected by the selection criteria 5 and 6.}  
    \label{fig:sideband_N-1_meanadc}
\end{figure}

 Mean ADC serves as the final discriminating variable for identifying monopole signals in the data. Since monopoles deposit significantly more energy in the detector than cosmic rays, they are expected to exhibit higher mean ADC values. Figure~\ref{fig:sideband_N-1_meanadc} shows the distribution of mean ADC, with all other cuts applied except for the one on mean ADC itself. The selection cuts 5 and 6 eliminate most irregular or clustered cosmic activity, leaving only a small residual background. Based on this distribution, different mean ADC thresholds were chosen for high- and low-$\beta$ monopoles to preserve efficiency throughout the velocity spectrum. High-$\beta$ monopoles, being highly ionizing, allow for a tighter mean ADC cut, while lower-$\beta$ monopoles, potentially less ionizing, require a lower threshold. After analyzing several $\beta$ points, we found that setting the relaxed mean ADC cut for $\beta < 0.03$ results in minimal loss of efficiency for lower-$\beta$ monopoles. The mean ADC selection requirement is the following:
    
 \begin{enumerate}
   \item[7.] {Mean ADC} $\geq 2500$  for $\beta_{reco}\geq 0.03$; 
   \item[8.] {Mean ADC} $\geq 2000$  for $\beta_{reco} < 0.03$.
 \end{enumerate}

Two effects degrade the selection efficiency at high monopole speeds. First, the delta rays can make the monopole track appear wider. This effect becomes relevant at $\beta \approx 0.68$ and rapidly worsens at higher speeds. To study the impact of these large delta rays, we relied on the custom simulation introduced in Sec.~\ref{sec:simulation}. Second, monopoles above the Cherenkov threshold produce a very large amount of light. In the majority of the detector, this produces no complications. However, at the detector edges where light is routed to the APDs, the WLS fibers of each set of 32 cells are routed through a common optical volume filled with scintillator. This volume is made of black plastic, which reflects little scintillation light, intended to suppress the collection of light produced in this volume, but monopole Cherenkov light is so bright that it can significantly illuminate some or all of the fibers in this region, making spurious hits at track ends. Similarly, at the far end from the APDs, there is a narrow channel connecting adjacent detector cells, which was necessary for filling the detector with scintillator. Under ordinary conditions, the light leakage through this channel is negligible, but monopole Cherenkov light can reflect through these channels sufficiently to produce hits in cells adjacent to the monopole path. We collectively refer to these effects as optical crosstalk. Because it degrades the apparent straightness and narrowness of monopole tracks, optical crosstalk causes a step down in selection efficiency at the Cherenkov threshold, $\beta \approx 0.68$.

We found that the  decrease of the trigger efficiency above Cherenkov threshold is due to crosstalk, while the decrease in overall efficiency above Cherenkov threshold arises from the combined influence of crosstalk and delta rays, which cause tracks to appear wider, as shown in Fig.~\ref{fig:trig_eff}. The drop in overall efficiency at low $\beta$ is primarily due to the mean ADC requirement for the signal.

 The dominant background in this analysis arises from the abundant atmospheric muons produced by cosmic-ray interactions. These are suppressed using event characteristics, as muon tracks are much less ionizing than the bright monopole signature. In addition, atmospheric muons often produce secondary showers, resulting in bulky and fragmented tracks. In contrast, a magnetic monopole makes a single, straight, and narrow track with uniform brightness along its path.

 A potential background for highly ionizing monopole candidates comes from cosmogenic muon bundles that traverse the detector nearly simultaneously and in parallel. Such closely spaced tracks could overlap and appear as a single highly ionizing event. To estimate this contribution, we simulated various muon bundle scenarios, including 2-, 3-, 4-, and 5-muon bundles. In each bundle, muons were generated entering the detector at the same position and time with the same angle. Additional samples of 2-muon bundles were also produced in which the muons had identical entry positions and angles, but were separated by a finite time offset between their entries into the detector. No bundles passed the selection cuts. To help estimate their potential contribution in the signal region, the selection cuts were relaxed. Still, none of the 2- or 3-muon bundle events passed the relaxed selection criteria; however, a few 4- and 5-muon bundle events did.
 
We searched for such possible 4-muon and 5-muon bundles in the minimum bias cosmic data collected by the FD. Muons within a bundle are expected to enter the detector with random spatial separations on the order of 100\,cm, consistent with their production in extensive air showers and subsequent multiple scattering. In data corresponding to an effective live time of four hours, ten 4-muon bundles were found with entry points within 100\,cm of one another, while no 5-muon bundles were observed. For purposes of this study, an $N$-muon bundle means that $N$ tracks were reconstructed. Muons that enter very close to each other will not be separately reconstructed; this is taken into account in the study. Assuming that multiple scattering gives rise to similar events with the muons in random positions within 100\,cm of each other, a Monte Carlo simulation showed that $5\times10^{-7}$ of 4-muon bundles and $5\times10^{-8}$ of 5-muon bundles have entry points all within 5\,cm of each other, corresponding to an exposure of 2,713~days. The simulation further estimates the probability that such muons stay close enough together to be reconstructed as a single track along their entire length. From the data and simulation, the expected background in the signal region in 2,713~days of exposure is $8\times10^{-7}$ events for 4-muon bundles and $< 2\times10^{-7}$ events, at 90\% C.L., for 5-muon bundles. Higher multiplicity bundles are even less likely to overlap well enough to mimic a monopole. These are overestimates, as we could only select any bundles in the Monte Carlo with relaxed selection criteria.

The background contribution was independently estimated using the sideband region of the mean ADC distribution in the ionization-based triggered data. In this study, several selection cuts were relaxed to obtain sufficient statistics in the sideband region, and the resulting mean ADC distributions were well described by exponential functions. The fitted parameters were then used to extrapolate the background estimate into the signal region. This study was performed separately for $\beta \geq 0.03$ and $\beta < 0.03$, as each corresponds to a distinct signal region. For $\beta \geq 0.03$, selection criterion~6 was relaxed, resulting in an extrapolated background of $(7 \pm 1) \times 10^{-5}$ in the signal region. Similarly, for the $\beta < 0.03$ region, selection criterion~6 and several preselection cuts were relaxed, including the removal of $dE/dx$, through-going, and minimum hit requirements, along with a relaxation of the mean ADC threshold, resulting in an extrapolated background of $(3 \pm 1) \times 10^{-6}$ events. This estimate, which contains no assumptions about the origin of potential backgrounds, agrees with the muon-bundle study that the Standard Model background is negligible.

\section{Results}
\begin{figure}
    \centering
    \includegraphics[width=\columnwidth]{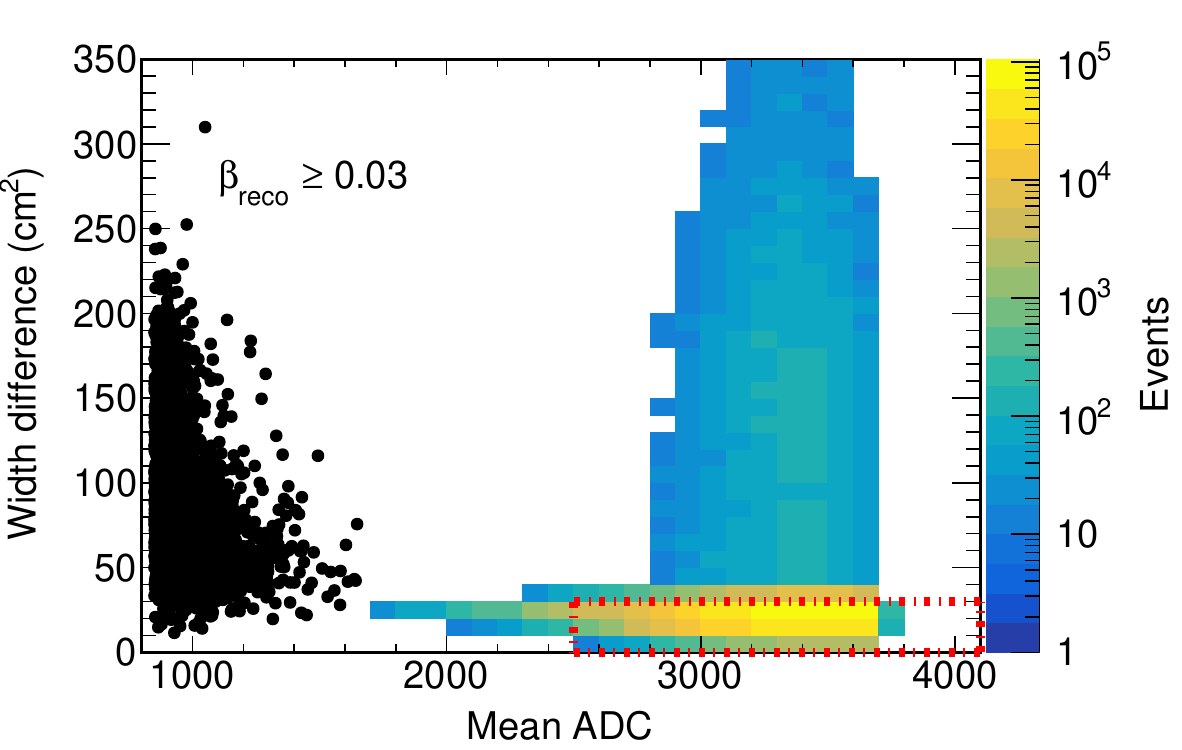}
    
    \includegraphics[width=\columnwidth]{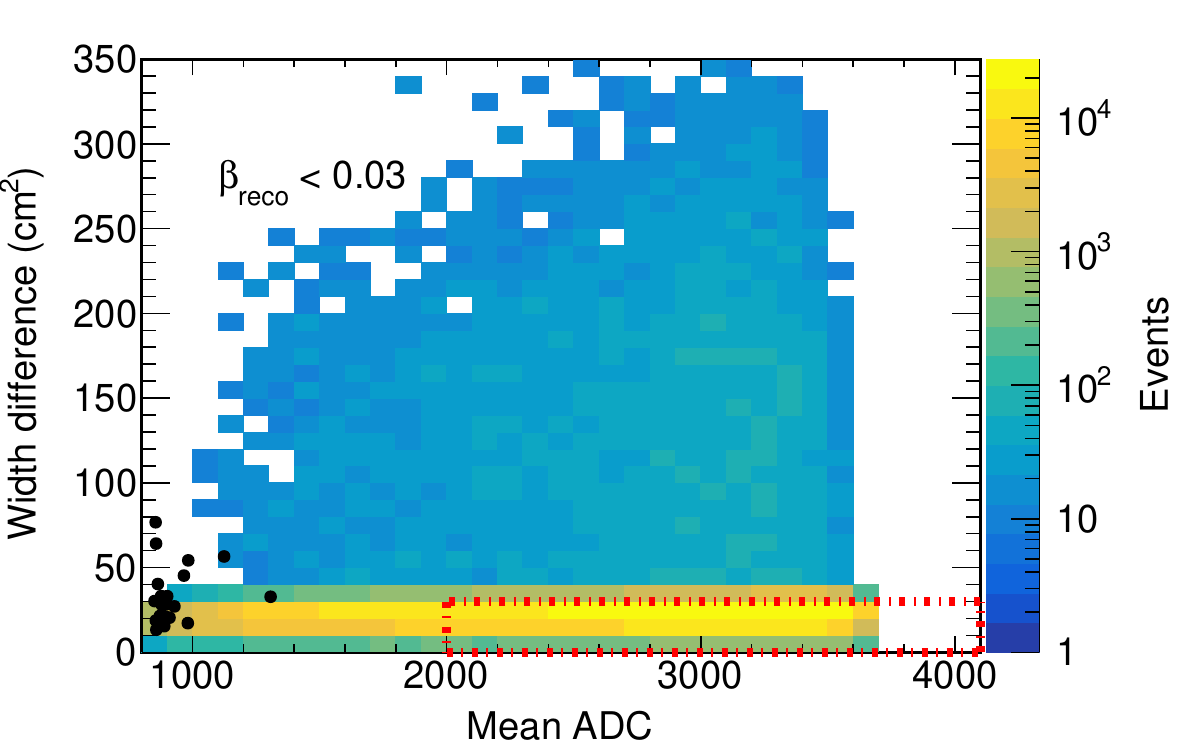}
    \caption{Distribution of width difference vs. mean ADC. Simulated signal events are shown as a colored heatmap, with data events shown as black points. The top panel corresponds to events with $\beta_{\text{reco}} \geq 0.03$, and bottom panel to $\beta_{\text{reco}} < 0.03$. The selection criteria 1--5, except for those on the plotted variables, have been applied. The dashed boxes mark the signal regions, in which no data events are observed (note suppressed zero on horizontal axis)}.
\label{fig:box_opened_result}
\end{figure} 

Once the selection criteria were finalized, the remaining 99.5\% of ionization-based triggered data (see Sec.~\ref{sec:offline_analysis}) was then analyzed, and no candidates were observed in the signal region. Figure~\ref{fig:box_opened_result} shows the distribution of the full data sample in width difference vs. mean ADC after application of all selection criteria except those on the two variables themselves. All events are far from the signal region. 

A visual inspection of the most signal-like events was performed both for the cases of $\beta \geq 0.03$ and $\beta < 0.03$. In the high-$\beta$ region, the most signal-like events are predominantly through-going muons that traverse the detector near its corner, resulting in a track that just passes the 5\,m threshold. In the low-$\beta$ region, the most signal-like events typically arise from two tracks that occur at the same location but at slightly different times, leading to mismatched tracks between the XZ and YZ views. Future improvements to the analysis could be made by refining the selection criteria to reject mismatched tracks.
 
As no events were observed in the signal region, and the expected background is negligible, we set a 90\% C.L. upper limit on the monopole flux. The limits are obtained using the solid-angle-averaged acceptance, derived by integrating the projected area $A_{\rm proj}(\phi, \theta)$ and efficiency over the solid angle from which the monopoles can arrive. The former is the projected area of the FD in the direction $(\phi,\theta)$, where $\theta$ is the zenith angle and $\phi$ is the azimuthal angle. It is a purely geometrical quantity, ranging from $2.7 \times 10^{6}$\,cm$^2$ for monopoles entering along the $z$ axis, which see only the smallest detector faces, to $1.3 \times 10^{7}\,\text{cm}^2$ for monopoles entering in the direction of a detector diagonal.

The detection efficiency depends on both the monopole trajectory, through variations in track length and detector response, and the monopole speed $\beta$. The total efficiency is the product $\epsilon(\phi,\theta,\beta)=\epsilon(\phi,\theta)\epsilon(\beta)$, where  $\epsilon(\phi, \theta)$ is the efficiency as a function of angle, and $\epsilon(\beta)$ is the efficiency as a function of speed. The angular efficiency $\epsilon(\phi,\theta)$ is approximately uniform with a value between 0.6 and 0.8, except that it is much lower for angles at which particles cross few detector planes, such as near the zenith.

We define the solid-angle-averaged acceptance as:

\begin{multline}
\left\langle \epsilon A\right\rangle 
\equiv \frac{1}{\Omega}
\int_{0}^{2\pi} \int_{0}^{\theta_{\max}}
A_{\rm proj}(\phi,\theta)\,
\\
\times \epsilon(\phi,\theta) \epsilon(\beta) \sin\theta \, d\theta \, d\phi,
\label{eq:aprojeff}
\end{multline}
where $\Omega$ is the accessible solid angle (see Sec.~\ref{sec:simulation}), and $\theta_{\max} = \arccos\!\left(1 - \frac{\Omega}{2\pi}\right)$. The limits of $\phi$ and $\theta$ are well-defined in this way as all trajectories incoming to the detector centered around the azimuth. The upper flux limit at 90\% C.L., with zero background, is given by   
\begin{equation}
    \phi_{90\%} = \frac{\ln(10)}{\langle \epsilon A\rangle\, \Omega\, t},
\end{equation}
where $t$ is the total live time. The factor $\ln(10)$ corresponds to the 90\% C.L. upper limit for a Poisson process with zero observed counts.

\begin{figure}
    \centering
    \includegraphics[width=\columnwidth]{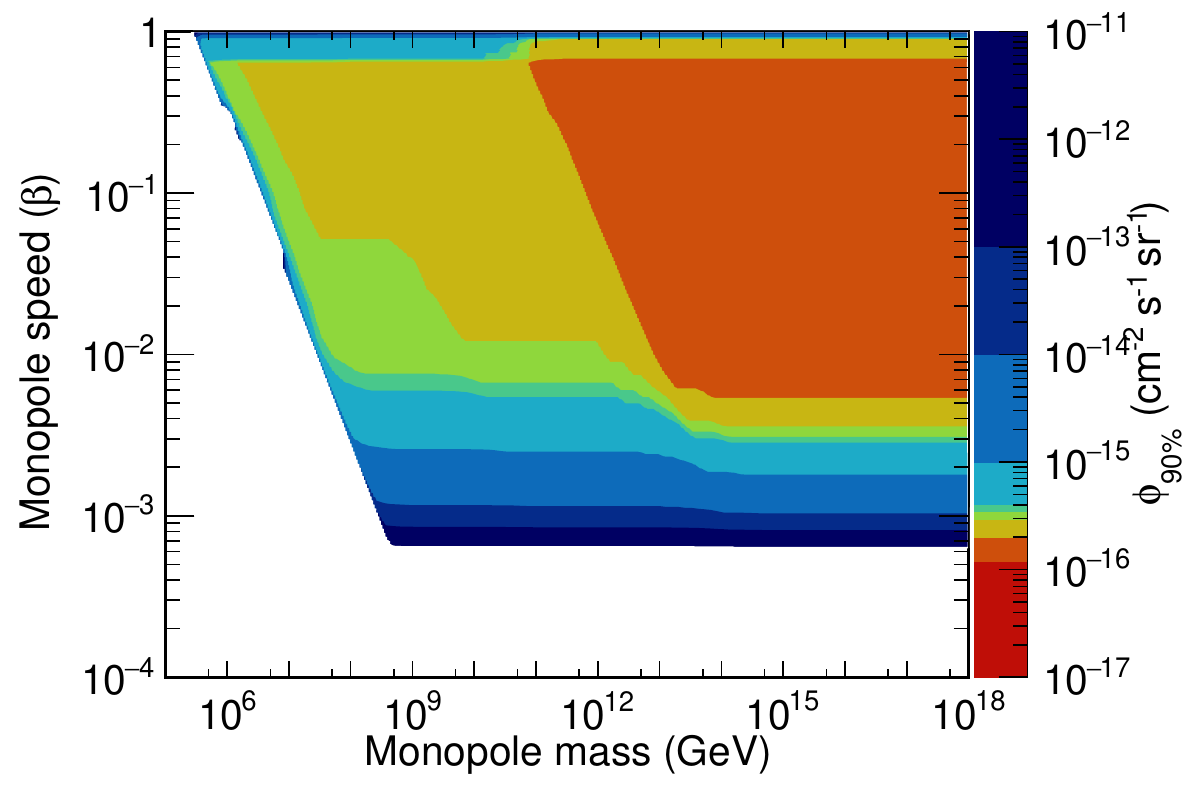}
    \caption{Flux limit as a function of monopole speed ($\beta$) and mass.}
\label{fig:flux-limit}
\end{figure}

\begin{figure}
    \centering
    \includegraphics[width=\columnwidth]{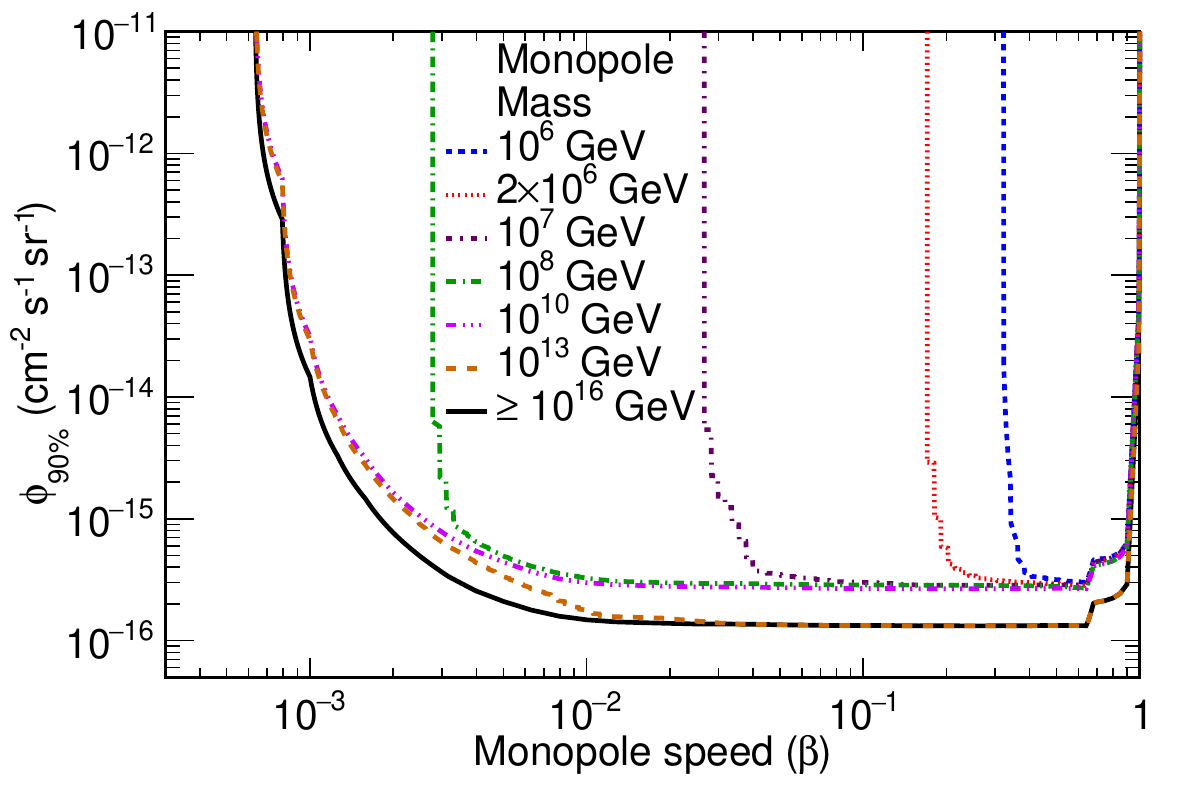}
    \caption{Flux limit vs. monopole speed ($\beta$) at fixed monopole masses. This plot is the graphical representation of the data in Tab. \ref{tab:flux_table}.}
    \label{fig:flux_slices}
\end{figure}

\begin{table}
\caption{90\% C.L. upper limits on the magnetic monopole flux in units of $10^{-16} \mathrm{ cm^{-2} s^{-1} sr^{-1}}$. \emph{Heavy} denotes monopoles with masses above $10^{16}~\mathrm{GeV}$ that are capable of traversing the entire Earth.}  
\centering
\begin{tabular}{l r r r r r}
\hline
\hline
$\log_{10}(\beta)$ & Heavy & $10^{13}\, \mathrm{GeV}$ &  $10^{10}\, \mathrm{GeV}$ & $10^7\, \mathrm{GeV}$ & $10^6\, \mathrm{GeV}$\\
\hline
$-3.1$ & 3000\phantom{.0} &  5800\phantom{.0}  &  6200\phantom{.0}   & --- &  --- \\
$-3.0$ & 150\phantom{.0} &   300\phantom{.0}   &  320\phantom{.0}    & --- &  --- \\
$-2.9$ & 33\phantom{.0} &    64\phantom{.0}    &  71\phantom{.0}   & --- &  --- \\
$-2.8$ & 15\phantom{.0} &    28\phantom{.0}    &  32\phantom{.0}   & --- &  --- \\
$-2.7$ & 7.8 &     15\phantom{.0}    &  17\phantom{.0} & --- & --- \\
$-2.6$ & 5.0 & 9.0 & 11\phantom{.0} & --- & --- \\
$-2.5$ & 3.4 & 6.0 & 7.3 & --- & --- \\
$-2.4$ & 2.6 & 4.4 & 5.4 & --- & --- \\
$-2.3$ & 2.1 & 3.4 & 4.4 & --- & --- \\
$-2.2$ & 1.8 & 2.7 & 3.7 & --- & --- \\
$-2.1$ & 1.6 & 2.1 & 3.2 & --- & --- \\
$-2.0$ & 1.5 & 1.7 & 3.0 & --- & --- \\
$-1.9$ & 1.4 & 1.6 & 2.9 & --- & --- \\
$-1.8$ & 1.4 & 1.6 & 2.9 & --- & --- \\
$-1.7$ & 1.4 & 1.5 & 2.8 & --- & --- \\
$-1.6$ & 1.4 & 1.4 & 2.8 & --- & --- \\
$-1.5$ & 1.4 & 1.4 & 2.8 & 16\phantom{.0} & ---  \\
$-1.4$ & 1.4 & 1.4 & 2.8 & 3.7 & ---  \\
$-1.3$ & 1.4 & 1.4 & 2.8 & 3.4 & ---  \\
$-1.2$ & 1.4 & 1.4 & 2.7 & 3.2 & ---  \\
$-1.1$ & 1.4 & 1.4 & 2.7 & 3.1 & ---  \\
$-1.0$ & 1.4 & 1.4 & 2.7 & 3.0 & ---  \\
$-0.9$ & 1.4 & 1.4 & 2.7 & 3.0 & ---  \\
$-0.8$ & 1.3 & 1.3 & 2.7 & 2.9 & ---  \\
$-0.7$ & 1.3 & 1.3 & 2.7 & 2.9 & ---  \\
$-0.6$ & 1.3 & 1.3 & 2.7 & 2.9 & ---  \\
$-0.5$ & 1.3 & 1.3 & 2.6 & 2.9 & ---  \\
$-0.4$ & 1.3 & 1.3 & 2.7 & 2.9 & 3.5  \\
$-0.3$ & 1.4 & 1.4 & 2.7 & 2.9 & 3.2  \\
$-0.2$ & 1.3 & 1.3 & 2.7 & 2.9 & 3.0  \\
$-0.15$ & 1.9 & 1.9 & 3.9 & 4.2 & 4.4  \\
$-0.1$ & 2.0 & 2.0 & 4.0 & 4.3 & 4.4  \\
$-0.05$ & 2.5& 2.5 & 5.0 & 5.3 & 5.4  \\
$-0.01$ &27\phantom{.0} & 27\phantom{.0} & 52\phantom{.0} & 57\phantom{.0} & 57\phantom{.0}  \\
$-0.004$ &  140\phantom{.0} & 140\phantom{.0}& 270\phantom{.0} & 300\phantom{.0} & 300\phantom{.0}   \\
$-0.0022$ & 760\phantom{.0} & 760\phantom{.0}& 1500\phantom{.0}	& 1600\phantom{.0} & 1600\phantom{.0} \\
                        
\hline 
\hline
\end{tabular}
\label{tab:flux_table}
\end{table}

  We compute the limit at each point in the space of monopole mass and monopole speed as shown in Fig.~\ref{fig:flux-limit}. Some representative flux limits are summarized in Tab.~\ref{tab:flux_table}, and Fig.~\ref{fig:flux_slices} provides a graphical illustration of these limits. The limits presented are conservative, as they use conservative estimates for the livetime and selection efficiency. These estimates include a 10\% reduction in the nominal $dE/dx$, as discussed in Sec.~\ref{sec:simulation}, to account for uncertainties in energy deposition. Additionally, a 10\% increase in the monopole energy loss in the Earth was applied when estimating the minimum mass required for a monopole to reach the FD at a given speed and incident angle. A 1\% reduction in live time was included to account for periods when incomplete information was available about the live time of the trigger. 

\section{Conclusion}
\begin{figure}
    \includegraphics[width=\columnwidth]{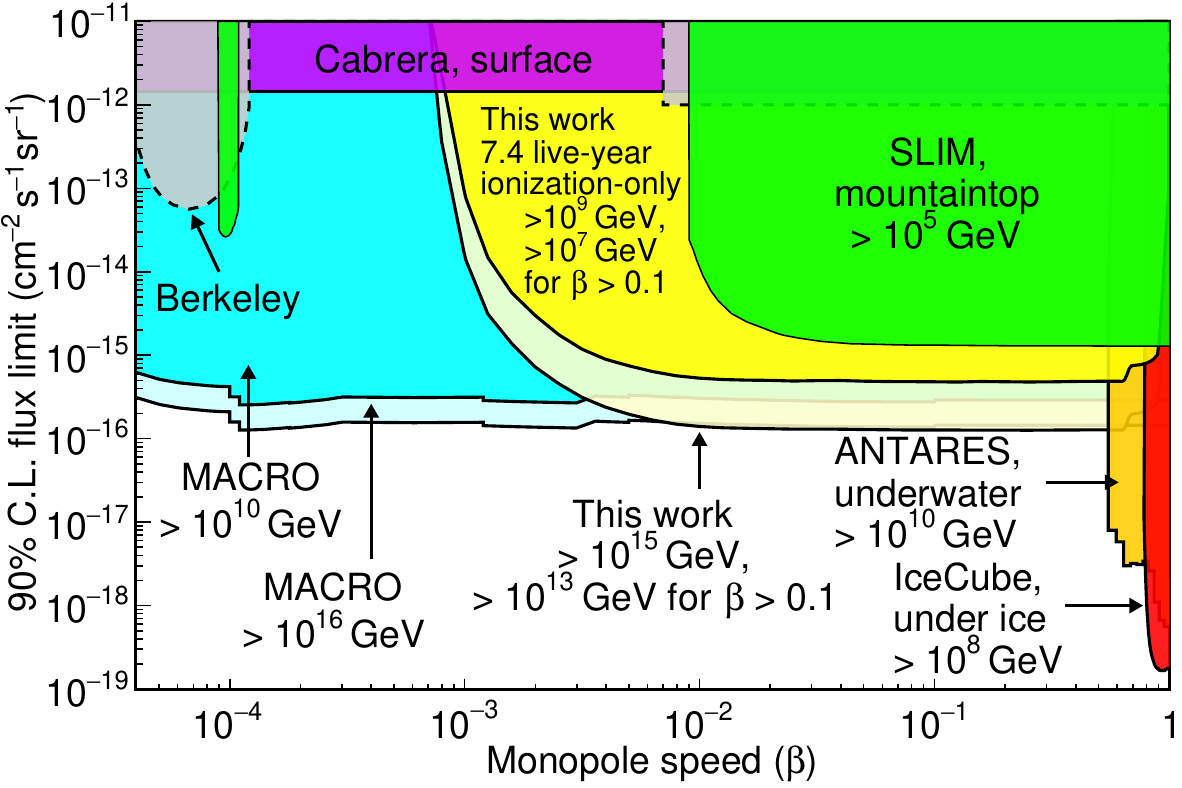}
    \caption{Upper limits on magnetic monopole flux for several experiments. In each region, the experiment with the strongest mass limit (least overburden) is shown above the others~\cite{cabrera1,slimresult,macroresult1,berkeley1,antares,antaressearchmagneticmonopolescomplete}. Results from our search are shown in the regions labeled ``This work.'' We report the most stringent limit, \mbox{$\phi_{90\%} < 8 \times 10^{-16}\,\mathrm{cm^{-2} s^{-1} sr^{-1}}$}, in the range \mbox{$0.005 < \beta < 0.8$}, corresponding to monopoles with masses greater than $10^{9}$ GeV for all speeds, and greater than $10^{7}$ GeV for $\beta > 0.1$. In the same range of $\beta$, for higher-mass monopoles capable of traversing the Earth, we set a flux limit  \mbox{$\phi_{90\%} < 2 \times 10^{-16}\,\mathrm{cm^{-2}s^{-1}sr^{-1}}$}, corresponding to monopoles with masses greater than $10^{15}$ GeV and $10^{13}$ GeV for $\beta > 0.1$.}
    \label{fig:flux-limit_comparison}
\end{figure}
 
We report limits as a function of monopole mass in the range \mbox{$7 \times 10^{-4} < \beta < 0.995$}. The limits presented in this paper are the strongest yet reported for some regions of speed and mass. The search was sensitive to monopoles with masses as low as \mbox{$2\times10^5\, \mathrm{GeV}$} for the fastest monopoles. For highly massive monopoles able to reach the detector from above or crossing the Earth from below, we report a flux limit \mbox{$\phi_{90\%} < 2 \times 10^{-16}\, \mathrm{ cm^{-2} s^{-1} sr^{-1}}$} at 90\% C.L. for monopoles with $0.005 < \beta < 0.8$.

Across the same range of speeds, we report a limit \mbox{$\phi_{90\%} < 8 \times 10^{-16}\, \mathrm{ cm^{-2} s^{-1} sr^{-1}}$} at 90\% C.L. for light  monopoles that can reach the detector through its minimal overburden. NOvA's combination of a large detector and minimal overburden enables sensitivity to slower and lower-mass monopoles not accessible in previous searches. This minimal overburden region is shown in Fig.~\ref{fig:flux-limit_comparison}, where NOvA's result is compared to previous experiments. The dependence on overburden is shown explicitly in the figure, where experiments with smaller overburden, and thus sensitivity to lower masses, are shown towards the top of the stack. The results presented here are complementary to our previous results using the transit time-based trigger, which cover a lower range of monopole speeds~\cite{slowmononova}. 

This search was sensitive to sub-relativistic highly-ionizing particles. Although we interpreted the absence of a signal in terms of a flux limit for magnetic monopoles, the data could be analyzed for many other hypothetical objects giving a similar signal, including quark nuggets~\cite{Witten_nugget1, nugget2, nugget3} and microscopic black holes~\cite{Arkani_Hamed_1998_black_hole1, Argyres_1998_black_hole2, Emparan_2000_black_hole3}. The interpretation of the data in these contexts is a subject for future work.

 \section{Acknowledgments}
This document was prepared by the NOvA Collaboration using the resources of the Fermi National Accelerator Laboratory (Fermilab), a U.S. Department of Energy, Office of Science, HEP User Facility. Fermilab is managed by the Fermi Forward Discovery Group, LLC, acting under Contract No. 89243024CSC000002. This work was supported by the U.S. Department of Energy; the U.S. National Science Foundation; the Department of Science and Technology, India; the European Research Council; the MSMT CR, GA UK, Czech Republic; the RAS, the Ministry of Science and Higher Education, and RFBR, Russia; CNPq and FAPEG, Brazil; UKRI, STFC and the Royal Society, United Kingdom; and the state and University of Minnesota. We are grateful for the contributions of the staffs of the University of Minnesota at the Ash River Laboratory, and of Fermilab. For the purpose of open access, the authors have applied a Creative Commons Attribution (CC BY) license to any Author Accepted Manuscript version arising.
 
\bibliography{FMM}

\end{document}